\documentclass[preprint,11pt]{aastex}               
\usepackage{epsfig}

\def\fun#1#2{\lower3.6pt\vbox{\baselineskip0pt\lineskip.9pt
  \ialign{$\mathsurround=0pt#1\hfil##\hfil$\crcr#2\crcr\sim\crcr}}}

\begin{document}

\title{Partial Field Opening and Current Sheet Formation in 
the Disk Magnetosphere}
\author{Dmitri A. Uzdensky}
\affil{Institute for Theoretical Physics, University of California} 
\affil{Santa Barbara, CA 93106}
\email{uzdensky@itp.ucsb.edu}
\date{\today}

\begin{abstract}
In this paper I analyze the process of formation of thin current structures 
in the magnetosphere of a conducting accretion disk in response to the 
field-line twisting brought about by the rotation of the disk relative to 
the central star. I consider an axisymmetric force-free magnetically-linked 
star--disk configuration and investigate the expansion of the poloidal field 
lines and partial field-line opening caused by the differential rotation 
between the star and a nonuniformly-rotating disk. I present a simple 
analytical model that describes the asymptotic behavior of the field in the 
strong-expansion limit. I demonstrate the existence of a finite (of order one
radian) critical twist angle, beyond which the poloidal field starts inflating
very rapidly. If the relative star--disk twist is enhanced locally, in some 
finite part of the disk (which may be the case for a Keplerian disk that 
extends inward significantly closer to the central star than the corotation 
radius), then, as the twist is increased by a finite amount, the field 
approaches a partially-open configuration, with some field lines going out to 
infinity. Simultaneous with this partial field opening, a very thin, radially
extended current layer forms, thus laying out a way towards reconnection in 
the disk magnetosphere. Reconnection, in turn, leads to a very interesting 
scenario for a quasi-periodic behavior of magnetically-linked star--disk 
systems with successive cycles of field inflation, opening, and reconnection. 
\end{abstract}



\section{Introduction}
\label{sec-intro}

\begin{figure}[tbp]
\begin{center}
\epsfig{file=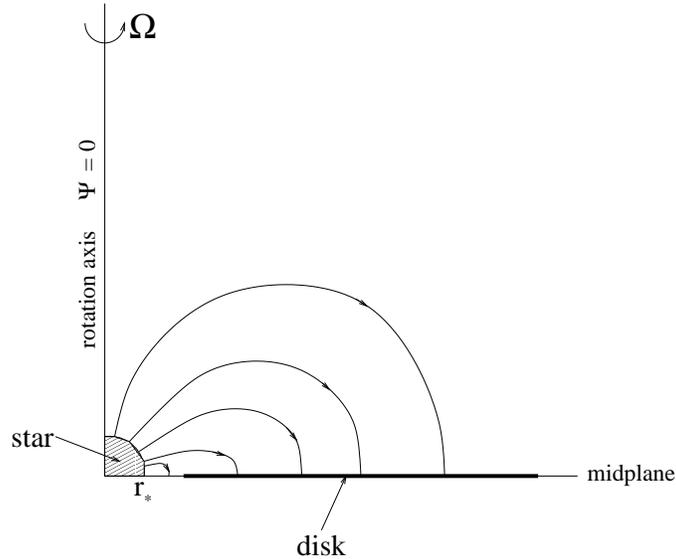,width=3.5 in}
\caption{Axisymmetric magnetically-linked star--disk system.
\label{fig-geometry}}
\end{center}
\end{figure}

Magnetic processes taking place in the magnetosphere above
a thin accretion disk play an important role in establishing 
the structure of disk outflows (winds and jets), in regulating 
the accretion flow, and in angular momentum transfer.
Of particular interest is the situation where there is a direct
magnetic connection between the star and the disk (the so-called
magnetically-linked star-disk system, see Fig.~\ref{fig-geometry}). 
This connection may lead to a direct angular momentum exchange between 
the disk and the star and is thus relevant for neutron star 
spin-up/spin-down events (Ghosh \& Lamb 1978, 1979; Wang 1987; 
Lovelace~et~al. 1995). It is also important because it provides 
a mechanism for direct channeling of accretion flow onto the polar 
regions of the star (Bertout, Basri, \& Bouvier 1988; Lamb 1989; 
Patterson 1994; K{\"o}nigl 1991).

\begin{figure}[tbp]
\begin{center}
\epsfig{file=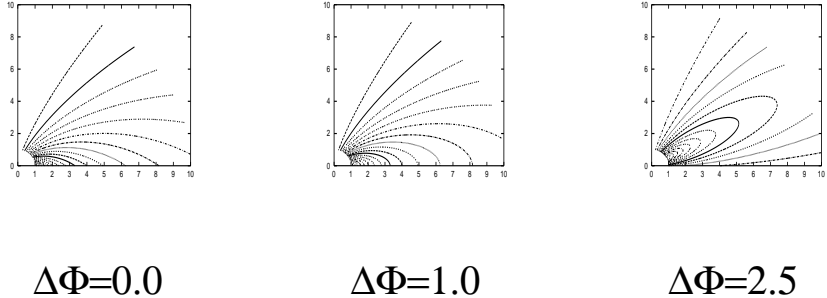,width=5 in}
\caption{Field-line expansion caused by the relative rotation
of the disk with respect to the star in the case of the $n=0.5$ 
self-similar model, where~$\Psi_{\rm disk}\sim r^{-n}$ (This 
figure is taken from Uzdensky et al. 2002a).
\label{fig-contour}}
\end{center}
\end{figure}

The evolution and even the sole existence of such a configuration
depends critically on the conducting properties of the disk, the
star, and the overlying magnetosphere. This complex physical problem 
can be studied on various time scales; one should start, however, 
with the shortest relevant time scale, namely, the rotation period. 
On this time scale both the central star and the disk can usually be 
approximated by ideal conductors (with the exception of the case when 
the central object is a black hole), and so can the low-density 
magnetosphere (or corona) that lies above the disk. In this case 
there is no steady state because the differential rotation between the 
disk and the star leads to continuous twisting of the field lines.
The magnetic field in the magnetosphere responds to this differential 
rotation  by rapid expansion driven by the increased toroidal field
pressure. The field lines become elongated along a direction making
a roughly $60^\circ$ angle with the rotation axis. This expansion 
process has been studied in some detail in the framework of the 
force-free model.%
\footnote{The validity of the force-free approximation in these systems
is justified by the fact that, due to very low density in the magnetosphere, 
the Alfv{\'e}n velocity there greatly exceeds both the sound speed and 
the rotation speed.}
In particular, semi-analytic self-similar models 
(van~Ballegooijen~1994, hereafter VB94; Lynden-Bell 
\& Boily 1994, hereafter LBB94; Uzdensky et al. 2002a)
have shown that the field-line expansion leads to an effective 
opening of the field lines after a finite (a few radians) twist 
angle (see Fig.~\ref{fig-contour}). This finite-time singularity 
has also been observed in numerically constructed non-self-similar 
sequences of force-free equilibria (Uzdensky et al. 2002a) and also 
in non-force-free full-MHD numerical simulations by Goodson et al. 
(1997, 1999). This process is essentially very similar to the process 
of finite-time field-line opening of a twisted coronal field studied 
extensively in solar physics (see, e.g., Barnes \& Sturrock 1972; 
Low 1986; Roumeliotis et al. 1994; Miki{\'c} \& Linker 1994; Wolfson 1995; 
Aly 1995). Thus, at present there is growing evidence that a finite-time 
opening of field lines in response to differential rotation is a generic 
feature of  force-free magnetospheres of magnetically-linked star--disk 
systems.

It is, however, still not clear what happens after this opening. 
Basically, one can envision two possible outcomes: the field lines 
may either reconnect (across the separatrix between the two domains 
with the opposite direction of the radial field) and close back; or 
they may stay open indefinitely. In the first scenario (Aly \& Kuijpers 
1990, VB94, Goodson et al. 1997, 1999) some resistivity is present in the 
magnetosphere, and as a result, a small amount of flux gets immediately 
reconnected, leading to the formation of an X-point. As reconnection 
progresses, a toroidal plasmoid forms in the magnetosphere. This 
plasmoid contains most of the toroidal flux; it becomes completely 
detached from both the disk or the star and just floats away,
presumably feeding a jet (Goodson et al. 1999). At the same time,
the magnetic link between the star and the disk is reestablished,
with the reconnected magnetic field lines being now twisted less
than they were just before the reconnection event. The poloidal 
magnetic field tension then quickly contracts them back towards 
the star, and the system returns to its initial state. Thus, in 
this scenario the situation is manifestly time-dependent; the time 
evolution consists of periodic cycles of successive field line 
inflation due to the differential shearing, ``effective''
field-line opening, reconnection, and the relaxation of the 
reconnected magnetic field to the initial state (of lesser
magnetic stress), which completes the cycle.

In the second scenario, investigated by Lovelace~et~al. (1995), 
there is no reconnection and a true steady state is established, 
whereby the field is at least partially open and the magnetic link 
between the star and a significant part of the disk is permanently 
broken.

The fact that the above two scenarios differ so dramatically 
raises the level of urgency of identifying the conditions 
under which each of them can be realized.
The main difference between these two pictures is that {\it 
magnetic reconnection} is allowed to take place in one but not in the
other. Thus, as it often happens in astrophysics and space
physics, the physics of magnetic reconnection plays a key role.
This issue is by no means trivial and it is questionable
whether it could be resolved even by a direct numerical 
simulation. Indeed, numerical simulations typically 
suffer from unrealistically large numerical resistivities
that make it very easy for oppositely directed field 
lines to reconnect (as seems to be the case with the simulations 
by Goodson et al. 1999). 

It is usually the case that in order to get efficient reconnection,
a thin current sheet is needed. In the two-dimensional axisymmetric 
situation discussed here, a natural place where such a current sheet 
can arise as the system approaches the open-field state is the current 
concentration region between the two domains with oppositely directed 
radial magnetic field. While the field lines are still closed, the 
current is concentrated along the apex line~$\theta=\theta_{\rm ap}(\Psi)$.%
\footnote
{Here, $\theta_{\rm ap}(\Psi)$ marks the angular position of the apex
of an inflated field line (an apex is defined as the farthest from the 
central star point on a field line). Usually, $\theta_{\rm ap}$ 
is close to $60^\circ$ and depends only very weakly on the field 
line~$\Psi$.}
As the field opens up, this apex line turns itself into the separatrix
between two regions of oppositely directed, open (i.e., extending to
infinity) field lines that comprise the open-field configuration.
If one now considers such a configuration (where the field lines 
have already been opened), then one discovers that the formation 
of a current sheet is essentially unavoidable, at least as long as 
the force-free approximation is valid.%
\footnote
{As was pointed out by the referee of this paper, 
if a heated atmosphere is present above the disk, 
then, once the magnetic field opens up, the drop 
of magnetic intensity with distance may be fast 
enough for the disk wind far out to dominate and 
keep the field from reconnecting. The centrifugal 
wind could also hold the field open provided there 
is a lower, sufficiently heated atmosphere to feed 
that wind. This may be the implied reasoning behind 
the apparent neglect of the possibility of reconnection 
by Lovelace's et al. (1995).}
Indeed, an {\it open} magnetic field is potential, it has no toroidal field. 
This is because all toroidal flux created by the initial twisting has 
escaped to infinity as the result of magnetic field expansion and 
opening. The field becomes predominantly radial everywhere, with $B_r$ 
reversing across the separatrix $\theta=\theta_{\rm ap}$. In the 
absence of the toroidal field $B_\phi$ at $\theta=\theta_{\rm ap}$, 
the pressure balance across the separatrix cannot be maintained, and 
this leads to the collapse of the magnetic configuration to one
with an infinitesimally thin current sheet; the two oppositely 
directed magnetic fields move toward each other, forming a thin 
current layer. Finally, the collapse is stopped when resistive 
(or other non-ideal) effects become important in the layer. Because 
the current density is tremendously increased, magnetic reconnection 
can occur.

It is important to realize, however, that, strictly speaking, 
as long as one has a force-free magnetic field with the field 
lines {\it closed}, one can never have a true, infinitesimally 
thin current sheet. Indeed, while the system is going through 
a sequence of equilibria of {\it closed} field lines, i.e., for 
$t < t_c$ (where $t_c$ marks the moment time of field-line opening), 
there is always finite toroidal field, $B_\phi$, present. 
This is because it is the toroidal flux that drives the expansion 
along $\theta=\theta_{\rm ap}$, and in a force-free equilibrium its outward 
pressure balances against the poloidal field's large curvature. Thus, 
in a closed force-free configuration the toroidal flux is kept in 
place by the poloidal field tension and prevents poloidal field 
lines from contracting back to a less-stressed state. At the same 
time, this toroidal field provides the pressure in the $\theta$~direction, 
which prevents the poloidal field collapse into a current sheet 
configuration.

One could ask, however, whether a current sheet can form {\it
asymptotically}, that is, whether the sequence of force-free 
equilibria, which governs the system's evolution, can asymptotically
lead to stronger and stronger thinning of the current concentration 
region as one approaches the critical moment, so that the 
characteristic angular width of this region, $\Delta\theta$,
goes to zero as $t\rightarrow t_c$. Note that this does not 
always happen. For example, in the self-similar force-free 
model for a uniformly rotating disk (VB94; Uzdensky et al. 2002a), 
where all magnetic quantities are power laws of~$r$ with fixed power-law 
indices, $\Delta\theta$ does not go to zero but approaches a 
finite value that depends solely on the flux distribution 
$\Psi_{\rm disk}(r)$ on the disk surface. In particular, if 
$\Psi_{\rm disk}(r) \sim r^{-n}$, then $\Delta\theta$ approaches 
a finite value of order a fraction of one radian if $n=O(1)$ (and
is proportional to~$n$ in the limit $n\rightarrow 0$), as the critical 
moment $t_c$ is approached. 

Asymptotic current sheet formation was in fact observed in 
the self-similar model by LBB94 (in cylindrical geometry) 
and also in an essentially very similar work by Wolfson 
(1995) (in spherical geometry in the solar corona context). 
This is explained by the fact that the power law index~$n$ 
was forced to change during the evolution. In particular, 
in the cylindrical case considered by LBB94, all magnetic 
flux was required to go through the boundary at some small 
radius (compared with the radius under consideration). 
Therefore, for each value of the power exponent~$n$, the 
resulting equilibrium corresponded to a member of the VB94 
family of solutions in the limit of infinite expansion, 
$t\rightarrow t_c(n)$ (this limit was analyzed in detail 
by Uzdensky et al. 2002a). Since in the VB94 analysis $t_c(n)$ 
grows as $n\rightarrow 0$, the gradual increase of the twist 
angle with time meant that, when considering the LBB94 solutions,
one were bound to obtain a sequence of solutions with ever diminishing 
values of~$n$ [determined from the condition $t_c(n)=t$]. That is, in 
the LBB94 sequence of solutions, $n$ had to decrease and approached 
zero at some point, which lead to the reported thinning and formation 
of a current sheet. Related to this change of~$n$ was the fact that 
the field-line footpoints were allowed (and actually had) to move 
poloidally. If, instead, one insisted on having the footpoints tied 
firmly to the disk surface, then the flux distribution on this surface 
(which, in the self-similar model, must be a power law extending to 
arbitrary large radii ) could not change, at least on the short, 
rotation-period time scale considered here. Then $n$ and, hence, 
$\Delta\theta$ would both stay finite. Under these circumstances 
the prospects for magnetic reconnection to occur in a timely 
manner would be very slim, at least in the two-dimensional 
framework (see Uzdensky et al. 2002b).

Thus, the transformation of the current concentration region 
into a true, infinitesimally-thin current sheet, even in the 
asymptotic sense, cannot, in general, be taken for granted. 

The failure of the self-similar force-free model to provide a plausible
current-sheet formation (and hence reconnection) scenario makes it
increasingly important to try to understand the asymptotic 
current-sheet formation 
process in a more general situation where the self-similar model does not 
apply. In particular, it is interesting to ask whether (still within
a force-free framework) the current-concentration region can
become very thin, thus indicating a way towards forming a true current 
sheet. If such a thin current layer does form, what is its structure?
In particular, what is the $\Psi$-profile of the angular width
$\Delta\theta$ and how does it depend on the twist angle profile
$\Delta\Phi(\Psi)$?

It is important to realize that in the self-similar model there 
is no special radial scale and hence all the field lines must open 
up simultaneously. In a general, non-self-similar situation, however, 
there is a possibility of a {\it partial field-line opening},%
\footnote
{The importance of partial field-line opening in solar coronal 
processes has been recognized and emphasized by Low (1990) and 
by Wolfson \& Low (1992), who suggested that a partially open 
field configuration may be energetically accessible even when 
a completely open configuration is not.}
with the domain of open field lines being adjacent to that of 
closed field lines. As we shall discuss below, this can greatly 
facilitate the current-sheet formation process.

On the intuitive level, the basic idea of how a thin 
magnetic structure can form in the disk magnetosphere is
simple. Imagine a {\it non-uniformly} rotating disk.
In particular, consider a system on the brink of the
partial field-line opening: let there be a field line
$\Psi_c$ such that, as the system approaches a certain
critical time $t_c$, the field lines outside of $\Psi_c$
(i.e., $\Psi<\Psi_c$) tend to open, while those inside
stay closed.\footnote
{We count poloidal magnetic flux $\Psi$ from the outside inward, 
i.e., we set $\Psi(\pi/2,r\rightarrow \infty) \rightarrow 0$.}
Now, if $\Delta\Omega(\Psi)$ [and hence the twist $\Delta\Phi(\Psi)=
\Delta\Omega t$] is nonuniform, say, rising outward, then the 
degree of expansion of field lines will also be non-uniform. 
Indeed, two neighboring field lines $\Psi_1=\Psi_c+\delta\Psi$ 
and $\Psi_2=\Psi_c-\delta\Psi$ will have their footpoints close 
to each other, while their apexes $r_{\rm ap}(\Psi_1)$ and 
$r_{\rm ap}(\Psi_2)$ will be very far apart since 
$r_{\rm ap}(\Psi_1)$ should stay finite whereas 
$r_{\rm ap}(\Psi_2) \rightarrow \infty$ as $t\rightarrow t_c$.
In other words, $d\log r_{\rm ap}(\Psi)/d\log\Psi$ at $\Psi=\Psi_c$
will go to infinity as $t\rightarrow t_c$.
As we shall show in this paper, 
a thinning of the apex region (i.e., current layer
formation) is characteristic for a situation like this,
which is consistent with the spirit of the studies by LBB94 
and by Wolfson (1995) who both show that a current sheet forms 
asymptotically when $n\equiv d\log\Psi/d\log r \rightarrow 0$.

Thus, the main thrust of this paper is to demonstrate
how twisting of field lines leads to the formation of
thin current structures that asymptotically become 
thinner and thinner as the field approaches a partially-open
state in finite time. In \S~\ref{sec-model} we present the 
basic geometry of the problem and a description of our model. 
In \S\S~\ref{sec-twist}--\ref{sec-radial-balance} we discuss 
the three main components of the model: \S~\ref{sec-twist} 
focuses on the purely geometrical relationship between the twist 
angle of the field lines and the relative strength of the toroidal 
field, while \S\S~\ref{sec-theta-balance} and~\ref{sec-radial-balance} 
describe our treatment of the~$\theta-$ and the radial force balance 
equations, respectively. In \S~\ref{sec-analysis} we formulate and discuss 
the final set of differential equations describing the behavior of the 
angular thickness of the current concentration region. Also in that 
section we consider three specific examples: \S~\ref{subsec-const-u} 
deals with the constant-twist case, \S~\ref{subsec-critical-twist} 
investigates the behavior in the vicinity of the critical twist
angle, and \S~\ref{subsec-numerical} describes a numerical solution 
of our equations for the case of a locally-enhanced twist angle.
We present our conclusions in \S~\ref{sec-conclusions}.

Finally, we note that although we consider this problem 
in the accretion disk context, our methods and the main 
findings can be directly applied in the solar corona context. 
In particular, we note that the differential-rotation-driven
process of partial field-line opening (and the detachment of
the toroidal plasmoid associated with it) is essentially very
similar to a coronal mass ejection; in addition, the partially 
opened field configuration considered in this paper is similar
to a coronal helmet streamer configuration (see, for example,
Low 2001).


\section{Description of the Model}
\label{sec-model}

In spherical coordinates $(r,\theta,\phi)$, the poloidal components of 
an axisymmetric magnetic field can be written in terms of the poloidal 
magnetic flux function~$\Psi$ as 
\begin{equation}
B_r = {1\over{r^2\sin{\theta}}} 
{{\partial\Psi}\over{\partial\theta}}, 
\label{eq-2-Br}
\end{equation}
\begin{equation}
B_\theta = -{1\over{r\sin{\theta}}} 
{{\partial\Psi}\over{\partial r}}. 
\label{eq-2-Btheta}
\end{equation}

In a force-free equilibrium the magnetic flux function satisfies
the Grad--Shafranov equation,
\begin{equation}
{{\partial^2\Psi}\over{\partial r^2}}+
{{\sin\theta}\over{r^2}} {\partial\over{\partial\theta}}
\left({1\over{\sin\theta}} {{\partial\Psi}\over{\partial\theta}}\right)=
-F F'(\Psi),
\label{eq-2-GS}
\end{equation}
where the {\it generating function}
\begin{equation}
F = B_\phi r \sin{\theta}
\label{eq-2-def-F}
\end{equation}
is $2/c$ times the total poloidal current flowing through the 
circle defined by $\phi\in [0,2\pi]$ at fixed $(r,\theta)$. 
Since in an axisymmetric equilibrium poloidal current must 
follow poloidal field lines, $F$ is constant along the field 
lines, i.e., $F=F(\Psi)$. 

\begin{figure}[tbp]
\begin{center}
\epsfig{file=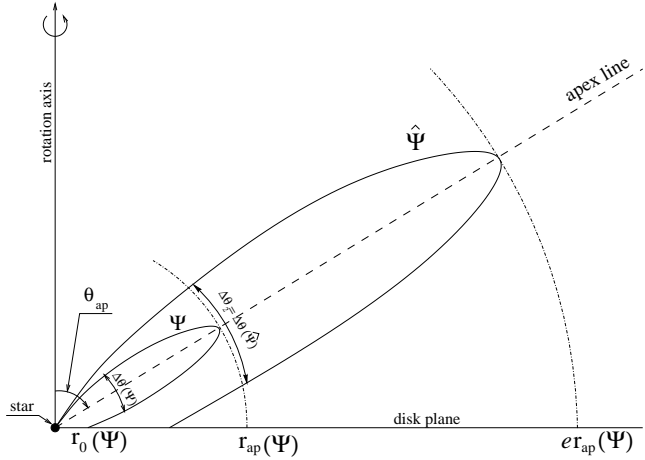,width=8.5cm}
\caption{Geometry of the problem: strongly expanded field lines
in the magnetosphere of a magnetically-linked star--disk system.
\label{fig-model}}
\end{center}
\end{figure}

Now, consider a configuration with the dipole-like field topology, and 
assume that the magnetic axis coincides with the rotation axes of both 
the disk and the star. We are interested in a situation where the system 
is approaching the point of field-line opening, so that the field lines 
are expanded very strongly along the direction of their apexes $\theta=
\theta_{\rm ap}$, so that $r_{\rm ap}(\Psi)\gg r_0(\Psi)$ (where 
$r_{\rm ap}(\Psi)$ and $r_0(\Psi)$ are the radial positions of the 
apex and the disk footpoint of field line $\Psi$; see Fig.~\ref
{fig-model}). In this asymptotic regime $\theta_{\rm ap}(\Psi)$ should not 
change much from one field line to another. In the self-similar model, for 
example, $\theta_{\rm ap}(\Psi)$ is, of course, exactly the same for all 
field lines and varies by only a few degrees as~$n$ changes from~0 to~1; 
and in the non-self-similar numerical model of Uzdensky et al. (2002a), 
$\theta_{\rm ap}(\Psi)$ lies between 60 and $70^\circ$. In any case, 
the $\Psi$~dependence of $\theta_{\rm ap}$ is much weaker than that 
of the field line parameters $r_{\rm ap}$ and~$\Delta\theta$ (the 
latter is defined immediately below); therefore, we shall neglect 
this dependence in our analysis and assume that the apex direction 
is the same for all field lines,
\begin{equation}
\theta_{\rm ap}(\Psi) = {\rm const}.  
\label{eq-2-apex-angle}
\end{equation}

Next, for a given field line $\Psi$, let us define the angular
width $\Delta\theta_1=\Delta\theta(\Psi)$ of the region enveloped 
by this field line. For definiteness, we define $\Delta\theta(\Psi)$ 
in terms of the shape $r(\theta,\Psi)$ of the field line as
\begin{equation}
r\left(\Psi,\theta_{\rm ap}+{\Delta\theta(\Psi)\over 2}\right) 
\equiv {{r_{\rm ap}(\Psi)}\over e} \simeq
r\left(\Psi,\theta_{\rm ap}-{\Delta\theta(\Psi)\over 2}\right)\, , 
\label{eq-2-def-deltatheta}
\end{equation}
where $e=2.7183...$ is the base of the natural logarithm.

In addition to $\Delta\theta_1=\Delta\theta(\Psi)$, we also define 
$\Delta\theta_2$ as $\Delta\theta_2(\Psi)\equiv \Delta\theta(\hat{\Psi})$, 
where $\hat{\Psi}$ is the field line with the apex radius~$r_{\rm ap}
(\hat{\Psi})$ equal to~$e$ times~$r_{\rm ap}(\Psi)$ (see Fig.~\ref
{fig-model}). The quantity~$\Delta\theta_2(\Psi)$ can also serve as 
an estimate for the characteristic angular width [at a given $r=
r_{\rm ap}(\Psi)$] of the region where most of the toroidal flux, 
as well as most of the toroidal current, are concentrated.

We are interested in the situation where thin structures 
are about to form in the magnetosphere. Therefore, throughout 
this paper we shall systematically assume that
\begin{equation}
\Delta\theta(\Psi) \ll 1,
\end{equation}
for all field lines~$\Psi$ under consideration.

Our main goal is to evaluate $\Delta\theta(\Psi)$ and to 
determine how it changes in response to an increased twist angle.

Our program has three main ingredients. First, there
is a purely geometric constraint relating the twist angle
$\Delta\Phi(\Psi)$ to the function $F(\Psi)$ and to the poloidal
field $B_\theta$. The other two ingredients are dynamic
--- they reflect the force-free magnetic force balance
in two directions, across and along the apex line
(the $\theta$ and the radial directions, respectively).
Upon the completion of this program we shall obtain three
equations relating three functions, $F(\Psi)$, $\Delta\theta(\Psi)$, 
and~$r_{\rm ap}(\Psi)$, to each other. This will allow us to 
investigate under what circumstances thin structures can form.
We start with the geometric constraint.

\section{The Twist Angle Constraint}
\label{sec-twist}

The twist angle $\Delta\Phi(\Psi)$ of a field line is defined
as the difference between the toroidal positions of the footpoints
of this field line on the surface of the disk and that of the star.
It can be written as 
\begin{equation} 
\Delta\Phi(\Psi) = F(\Psi) I(\Psi).
\label{eq-3-twist}
\end{equation}

Here, the function $I(\Psi)$ is given by an integral along 
the field line~$\Psi$:
\begin{equation}
I(\Psi)= \int\limits_\Psi {{d\theta}\over{B_\theta r \sin^2{\theta}}}. 
\label{eq-3-def-I}
\end{equation}

Using equation~(\ref{eq-2-Btheta}), we can express the poloidal 
field~$B_\theta$ in terms of the function~$r(\Psi,\theta)$ as
\begin{equation}
B_\theta = -{1\over{r\sin{\theta}}} 
{{\partial\Psi}\over{\partial r}} =  -{1\over{r\sin{\theta}}}
\left({{\partial r}\over{\partial\Psi}}\right)^{-1}.
\end{equation}

Then,
\begin{equation}
I(\Psi)=-\int\limits_{\theta_*}^{\pi/2}
{{d\theta}\over{\sin\theta}}\, {{\partial r(\Psi,\theta)}\over{\partial\Psi}}
\simeq - {\partial\over{\partial\Psi}}
\int\limits_{\theta_*}^{\pi/2}
{{d\theta}\over{\sin\theta}}\, r(\Psi,\theta)\, , 
\label{eq-3-I-1}
\end{equation}
where we have neglected the dependence of $\theta_*$ (the angular 
position of the footpoint on the surface of the star) on~$\Psi$. 
We can do this because, in the regime of interest to our present 
study, only a relatively small amount of twist resides near the 
disk and the star surfaces, while the dominant contribution to 
the integral $I(\Psi)$ and, hence, to the total twist~$\Delta\Phi$ 
for a strongly expanded field line comes from the part of the field 
line near the apex, $\theta=\theta_{\rm ap}$. This is justified by 
the property of the twist to propagate along the flux tube and 
concentrate in the region of the weakest field (see Parker~1979), 
which in our case coincides with the apex region. 

Furthermore, we can see that, if a field line is expanded strongly, 
then we can write
\begin{equation}
\int\limits_{\theta_*}^{\pi/2} {{d\theta}\over{\sin\theta}}\, r(\Psi,\theta)
\equiv \kappa\, \Delta\theta(\Psi)\,  
{{r_{\rm ap}(\Psi)}\over{\sin\theta_{\rm ap}}}\, .
\label{eq-3-def-kappa}
\end{equation}

\begin{figure}[tbp]
\begin{center}
\epsfig{file=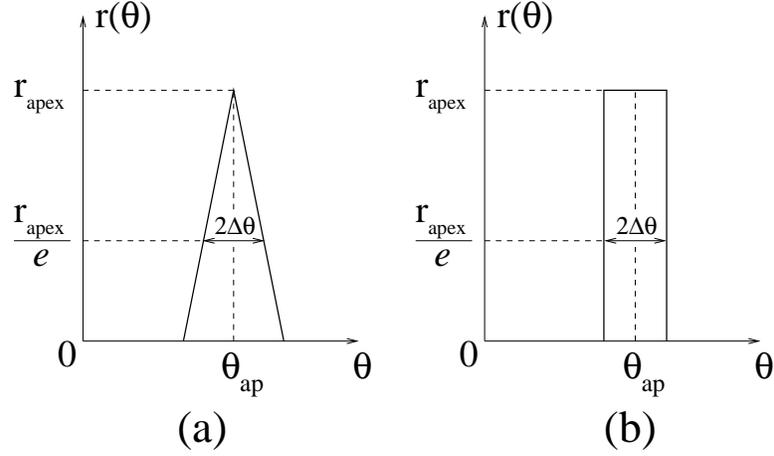,width=4 in}
\caption{The two examples of field-line shapes used in the text 
to demonstrate the calculation of the parameter~$\kappa$:
(a) the triangular shape; (b) the rectangular shape.
\label{fig-kappa}}
\end{center}
\end{figure}

This equation serves as the definition of the parameter $\kappa$.
Basically, this parameter describes the shape of the part of the
field line near the apex [i.e., the part giving the main contribution 
to~$I(\Psi)$], independent of the angular size $\Delta\theta(\Psi)$ 
and the radial size $r_{\rm ap}(\Psi)$, which appear in equation~(\ref
{eq-3-def-kappa}) explicitly. If all the field lines keep the same basic 
functional form near the apex, while only the typical angular and radial 
scales differ, we can regard $\kappa$ as being independent of $\Psi$. The 
validity of the assumption $\kappa={\rm const}$ is further strengthened 
by the observation that $\kappa$ varies very little over a broad range 
of field-line shapes. To demonstrate this, we here explicitly calculate, 
as an illustration, the values of~$\kappa$ for two extremely different 
shapes, the triangular shape (Fig.~\ref{fig-kappa}a) and the rectangular 
shape (Fig.~\ref{fig-kappa}b).

In the triangular case, shown in Figure~\ref{fig-kappa}a, 
we have, using the fact that $\Delta\theta \ll 1$ and hence
$\sin\theta \simeq \sin\theta_{\rm ap} = {\rm const}$,
$$
\int {{d\theta}\over{\sin\theta}} r(\Psi,\theta) \simeq 
{1\over 2} {{r_{\rm ap}(\Psi)}\over{\sin\theta_{\rm ap}}}
{{\Delta\theta}\over{1-1/e}},
$$
and so
$$
\kappa_a = {{e/2}\over{e-1}} \simeq 1.3 \, . 
$$

In the rectangular case, shown in Figure~\ref{fig-kappa}b, 
we again use $\Delta\theta \ll 1$ to approximate 
$\sin\theta \simeq \sin\theta_{\rm ap} = {\rm const}$,
and obtain
$$
\int {{d\theta}\over{\sin\theta}} r(\Psi,\theta) =
{{r_{\rm ap}(\Psi)}\over{\sin\theta_{\rm ap}}}\, \Delta\theta\, ,
$$
and so
$$
\kappa_b = 1.0 \, .
$$
Thus, from now on, we shall assume that $\kappa$ is constant.

Using equations~(\ref{eq-3-I-1}) and~(\ref{eq-3-def-kappa}) 
we can now write
\begin{equation}
I(\Psi)= - {\kappa\over{\sin\theta_{\rm ap}}} {d\over{d\Psi}} \, 
\Bigl( \Delta\theta r_{\rm ap} \Bigr) \, .
\label{eq-3-I-2}
\end{equation}

Substituting this into equation~(\ref{eq-3-twist}), we get
\begin{equation}
F(\Psi)=-{{\Delta\Phi(\Psi)}\over{\kappa r_{\rm ap}(\Psi)}} \, 
{{\sin\theta_{\rm ap}}\over{\Delta\theta(\Psi)}} \ 
{\Psi\over{(d\log \Delta\theta/d\log\Psi)-S(\Psi)}}\, ,
\label{eq-3-F}
\end{equation}
where 
\begin{equation}
S(\Psi) \equiv \biggl| {{d\log r_{\rm ap}(\Psi)}\over {d\log\Psi}} \biggr| = 
-\, {\Psi\over{r_{\rm ap}}}\, {{dr_{\rm ap}}\over{d\Psi}}\, . 
\label{eq-3-def-S}
\end{equation}


\section{The $\theta$~Force Balance}
\label{sec-theta-balance}

Consider the $\theta$ component of the force balance equation:
\begin{eqnarray}
{4\pi\over c}[{\bf j\times B}]_\theta &=& 
{4\pi\over c} \Bigl( j_\phi B_r - j_r B_\phi \Bigr) \nonumber \\
&=& {B_r\over r} {\partial\over{\partial r}} (r B_\theta)\, - \,  
{{B_r}\over r} {{\partial B_r}\over{\partial\theta}}\, - \, 
{{B_\phi}\over{r\sin\theta}}\, {\partial\over{\partial\theta}}\, 
\Bigl(\sin\theta B_\phi \Bigr) = 0 \, . 
\label{eq-4-j_cross_B_theta}
\end{eqnarray}

Here, the first term represents the $\theta$~component
of the magnetic tension force, and the second and the 
third terms represent the $\theta$~derivatives of the 
radial and the toroidal field pressures, respectively.
Intuitively, it is clear that the first term must be 
small compared with the other two terms, because the 
primary force balance across the apex region should 
be between the toroidal field pressure at the apex
and the radial field pressure outside the apex. The 
magnetic tension acts mainly in the radial direction
while its $\theta$~component is small. The same conclusion
can also be reached upon noticing that the first two terms 
in equation~(\ref{eq-4-j_cross_B_theta}) come from the 
corresponding two terms in the expression for the the toroidal 
current density, $j_\phi=(1/r)(\partial/\partial r)(rB_\theta) 
-(1/r) (\partial B_r/\partial\theta)$. It is intuitively clear
that, in the configuration where the field lines are strongly
elongated in the radial direction and the radial field changes
its sign across the apex line over a very narrow region, the
second term gives a much greater contribution to $j_\phi$ than
the first term. In \S~\ref{sec-analysis}, we shall come back to 
this conclusion and make a more rigorous {\it a posteriori} check 
of its validity, but now we shall just neglect the first term in 
equation~(\ref{eq-4-j_cross_B_theta}) and thus get
\begin{equation}
{{B_\phi}\over{\sin\theta}} {\partial\over{\partial\theta}}
\Bigl(B_\phi \sin\theta \Bigr) +
B_r {{\partial B_r}\over{\partial\theta}} = 0. 
\label{eq-4-force_balance_theta}
\end{equation}

Notice that $B_\phi$ changes on the angular scale $\Delta\theta \ll 1$,
while $\sin\theta$ changes on the angular scale of order 1~rad. Therefore,
we can neglect the variation of $\sin\theta$ with $\theta$ and write
\begin{equation}
{\partial\over{\partial\theta}} \, 
\Bigl( B_\phi^2 + B_r^2 \Bigr) = 0 \qquad \Rightarrow \qquad 
B_\phi^2 + B_r^2 = {\rm const} \qquad {\rm at~fixed}\ r.
\end{equation}

At the apex $\theta=\theta_{\rm ap}$, $B_r=0$, while outside
the apex region, $\theta > \theta_{\rm ap} + \Delta\theta_2/2$, 
$B_\phi$ becomes small. Hence, we obtain (consistent with the 
intuitive explanation given above) the pressure balance between 
the toroidal field at the apex and the radial field pressure 
outside the apex:
\begin{equation}
\Bigl| B_\phi(r_{\rm ap},\theta_{\rm ap}) \Bigr| = 
\Bigl| B_r\left(r_{\rm ap},
\theta>\theta_{\rm ap}+{{\Delta\theta_2}\over 2}\right) \Bigr|
\label{eq-4-Bphi=Br}
\end{equation}

The toroidal field $B_\phi(r_{\rm ap},\theta_{\rm ap})$ is easily
expressed in terms of~$F(\Psi)$:
\begin{equation}
B_\phi(r_{\rm ap},\theta_{\rm ap}) =
{{F(\Psi)}\over{r_{\rm ap}(\Psi)\sin\theta_{\rm ap}}}.
\label{eq-4-Bphi}
\end{equation}

As for $B_r(r_{\rm ap},\theta>\theta_{\rm ap}+\Delta\theta_2/2)$,
we shall estimate it as follows. The total poloidal flux~$\Delta\Psi$ 
through the interval $\theta\in [\theta_{\rm ap}+\Delta\theta_2/2,\pi/2]$
per one radian in the toroidal direction at fixed $r=r_{\rm ap}(\Psi)$ is 
equal to $\Delta\Psi=\Psi(r,\theta_{\rm ap}+\Delta\theta_2/2)-\Psi(r,\pi/2)$.
But, since we assume that $\Delta\theta_2 \ll 1$, we can approximate 
$\Psi(r_{\rm ap}(\Psi),\theta_{\rm ap}+\Delta\theta_2/2) \approx 
\Psi(r_{\rm ap}(\Psi),\theta_{\rm ap})\equiv \Psi$. In addition, for 
strongly expanded field lines, $r_{\rm ap}(\Psi) \gg r_0(\Psi)$, and 
hence $\Psi \gg \Psi[r_0=r_{\rm ap}(\Psi)]$ (where we take~$\Psi=0$ at 
infinity). Thus, we can evaluate $\Delta\Psi$ as $\Delta\Psi\approx\Psi$.

On the other hand, $\Delta\Psi$ can be estimated as
\begin{equation}
\Delta\Psi = - r_{\rm ap}^2(\Psi) 
\int\limits_{\theta_{\rm ap}+\Delta\theta_2/2}^{\pi/2}
B_r \sin\theta \,  d\theta.
\end{equation}

In the interval under consideration (i.e., $\theta\in 
[\theta_{\rm ap}+\Delta\theta_2/2,\pi/2]$), $B_r$ is 
roughly uniform, and so, using $\Delta\Psi\approx\Psi$,
we can write
\begin{equation}
B_r\Bigl( r_{\rm ap},\theta>\theta_{\rm ap}+{{\Delta\theta_2}\over 2}\Bigr) 
\simeq -{{a_1 \Psi}\over{r_{\rm ap}^2(\Psi)}} < 0,
\label{eq-4-Br}
\end{equation}
where 
\begin{equation}
a_1^{-1} \equiv 
\int\limits_{\theta_{\rm ap}+\Delta\theta_2/2}^{\pi/2}
\sin\theta d\theta= 
\cos \Bigl( \theta_{\rm ap}+{{\Delta\theta_2}\over 2} \Bigr) \simeq
\cos \theta_{\rm ap} \simeq 2
\label{eq-4-def-a1}
\end{equation}
is a constant.

Note that estimate~(\ref{eq-4-Br}) is consistent with 
the self-similar model, which would give (see \S\S~2.1 
and~2.3 of Uzdensky et al. 2002a)
\begin{equation}
B_r^{\rm self-similar}\Bigl( r_{\rm ap},
\theta>\theta_{\rm ap}+{{\Delta\theta_2}\over 2}\Bigr)
\simeq {{G_0'(\pi/2)}\over{G_{0,\rm max} \sin\theta_{\rm ap}}} \, 
{\Psi\over{r_{\rm ap}^2(\Psi)}} \, .
\nonumber
\end{equation}

Since the function $G_0(\theta)$ is roughly linear in the interval
$[\theta_{\rm ap}+\Delta\theta_2/2,\pi/2]$, we can write 
$G_{0,\rm max} \simeq -G_0'(\pi/2) \left(\pi/2-\theta_{\rm ap}\right)$, 
and so 
\begin{equation}
B_r^{\rm self-similar} \Bigl( r_{\rm ap},
\theta>\theta_{\rm ap}+{{\Delta\theta_2}\over 2}\Bigr) 
\simeq -{\Psi\over{r_{\rm ap}^2(\Psi) \, \sin\theta_{\rm ap}}}
\, {1\over{\pi/2-\theta_{\rm ap}}} \simeq
-{{a_1\Psi}\over{r_{\rm ap}^2(\Psi)}}\, . 
\end{equation}

Thus, combining equations~(\ref{eq-4-Bphi=Br}), 
(\ref{eq-4-Bphi}), and~(\ref{eq-4-Br}), we get
\begin{equation}
F(\Psi)=a_1 \sin\theta_{\rm ap}\, {\Psi\over{r_{\rm ap}(\Psi)}}\, .
\label{eq-4-F}
\end{equation}
Combining this with equation~(\ref{eq-3-F}), we get
\begin{equation}
{{d\log\Delta\theta(\Psi)}\over{d\log\Psi}}-S(\Psi)=
-{{\Delta\Phi(\Psi)}\over{\kappa a_1 \Delta\theta(\Psi)}}\, .
\label{eq-4-twist_constraint}
\end{equation}
This equation presents a physically sound and adequate description of 
the system's behavior, with most of the uncertainties associated with 
its derivation being hidden in the constants~$\kappa$ and~$a_1$.


\section{The Radial Force Balance}
\label{sec-radial-balance}

Consider now the radial force balance along the apex line
$\theta=\theta_{\rm ap}$. 
Physically, we have here a balance between the radial outward pressure 
of the toroidal field, the term on the right hand side, and the poloidal 
field tension, second term on the left hand side of the Grad--Shafranov 
equation~(\ref{eq-2-GS}). The first term on the left hand side, the 
poloidal field pressure, is small in our limit $\Delta\theta \ll 1$. 
The fact that this first term is much smaller than the second term 
can also be understood as the statement that, in the expression for 
the toroidal current density, $j_\phi=(1/r)(\partial/\partial r)(rB_\theta)-
(1/r) (\partial B_r/\partial\theta)$,
the first term is much smaller than the second term.
Thus, we see that the criterion for validity of the 
assumption that the first term in equation~(\ref{eq-2-GS}) 
is much smaller than the second term coincides with the 
criterion for validity of a similar assumption we made 
when deriving equation~(\ref{eq-4-force_balance_theta}) 
from equation~(\ref{eq-4-j_cross_B_theta}). Thus, both 
these assumptions will be justified simultaneously in
the next section.

So, we now have
\begin{equation}
\biggl. {{\sin\theta_{\rm ap}}\over{r_{\rm ap}^2}}\, 
{\partial\over{\partial\theta}}\, 
\left({1\over{\sin\theta}} {{\partial\Psi}\over{\partial\theta}}\right)
\biggr|_{\rm apex}= -F F'(\Psi), .
\label{eq-5-radial_force_balance-1}
\end{equation}
(Note that our interpretation of the Grad--Shafranov
equation only makes sense if the toroidal field pressure
is directed outward, which corresponds to $dF^2/d\Psi>0$).

Neglecting, as usual, the $\theta$ variation of $\sin\theta$
in comparison with that of the magnetic field quantities, we 
write
\begin{equation}
\biggl. {1\over{r_{\rm ap}^2}} {{\partial^2\Psi}\over{\partial\theta^2}}
\biggr|_{\rm apex} = -F F'(\Psi), .
\label{eq-5-radial_force_balance-2}
\end{equation}

We can estimate $\partial^2\Psi/\partial\theta^2$ 
at $\theta=\theta_{\rm ap}$ as
\begin{equation}
\biggl. {{\partial^2\Psi}\over{\partial\theta^2}} \biggr|_{\rm apex}=
\biggl. a_2 {{(\partial\Psi/\partial\theta)
|_{\theta_{\rm ap}+\Delta\theta_2/2} -
(\partial\Psi/\partial\theta)|_{\theta_{\rm ap}}}
\over{\Delta\theta_2/2}} \biggr|_{r=r_{\rm ap}},
\end{equation}
where $a_2$ is another constant of order one.

At the apex $\partial\Psi/\partial\theta \equiv 0$ by definition, so
\begin{equation}
\biggl. {{\partial^2\Psi}\over{\partial\theta^2}} \biggr|_{\rm apex}=
\biggl. {{2a_2}\over{\Delta\theta_2}}\, 
{{\partial\Psi}\over{\partial\theta}}\,  
\biggr|_{\theta_{\rm ap}+\Delta\theta_2/2}=
{{2a_2}\over{\Delta\theta_2}} r_{\rm ap}^2 \sin\theta_{\rm ap}
B_r\left(r_{\rm ap},\theta=\theta_{\rm ap}+{{\Delta\theta_2}\over 2}\right).
\end{equation}

Using our estimate~(\ref{eq-4-Br}) for 
$B_r(r_{\rm ap},\theta=\theta_{\rm ap}+\Delta\theta_2/ 2)$, we get
\begin{equation}
\biggl. {{\partial^2\Psi}\over{\partial\theta^2}}\biggr|_{\rm apex}=
-{{2 a_1 a_2}\over{\Delta\theta_2}}\, \Psi\, \sin\theta_{\rm ap}\, .
\label{eq-5-d2Psi/dtheta2}
\end{equation}

Then, from equation~(\ref{eq-5-radial_force_balance-2}),
\begin{equation}
FF'(\Psi)={{2 a_1 a_2}\over{\Delta\theta_2}} {\Psi\over{r_{\rm ap}^2}}\, .
\end{equation}

Thus we have obtained the following system of equations:
\begin{eqnarray}
F(\Psi) &=& a_1 \sin\theta_{\rm ap} {\Psi\over{r_{\rm ap}(\Psi)}}\, ,
\label{eq-5-F} \\
FF'(\Psi) &=& {{2a_1 a_2}\over{\Delta\theta_2}}{\Psi\over{r_{\rm ap}^2}}\, ,
\label{eq-5-FF'} \\
{{d\log\Delta\theta(\Psi)}\over{d\log\Psi}}-S(\Psi) &=&
-{{\Delta\Phi(\Psi)}\over{\kappa a_1 \Delta\theta(\Psi)}}\, .
\label{eq-5-twist_constraint}
\end{eqnarray}

We view this system as a system of three equations for three
functions, $F(\Psi)$, $r_{\rm ap}(\Psi)$, and $\Delta\theta(\Psi)$.

To simplify this system, we shall first use equation~(\ref{eq-5-F}) 
to eliminate function~$F(\Psi)$. We have
\begin{equation}
F'(\Psi)={{a_1 \sin\theta_{\rm ap}}\over{r_{\rm ap}(\Psi)}}\, 
\Bigl[ 1+S(\Psi) \Bigr]\, .
\end{equation}

Substituting this, together with equation~(\ref{eq-5-F}), 
into equation~(\ref{eq-5-FF'}), we get
\begin{equation}
\Bigl[ 1+S(\Psi) \Bigr] \, \Delta\theta_2(\Psi)=
{{2a_2}\over{a_1\sin^2\theta_{\rm ap}}}\equiv C^{-1}={\rm const} = O(1)\, .
\label{eq-5-def-C}
\end{equation}
For example, for $a_1=1/2$, $a_2=1$, and~$\sin\theta_{\rm ap}\simeq 1$,
we have $C\simeq 1/4$.

Now note that, because we are in the small-$\Delta\theta$ regime,
from equation~(\ref{eq-5-def-C}) it follows that
\begin{equation}
S(\Psi) \equiv \biggl| {{d\log r_{\rm ap}}\over{d\log\Psi}} \biggr| \gg 1.
\label{eq-5-S<<1}
\end{equation}
Then equation~(\ref{eq-5-def-C}) becomes
\begin{equation}
C S(\Psi) \Delta\theta_2(\Psi) = 1 \, .
\label{eq-5-C_S_Deltatheta2}
\end{equation}

It is interesting to note that, since $S=-d\log r_{\rm ap}/d\log\Psi$, 
then $\Psi(r,\theta_{\rm ap})\sim r^{-1/S}$, that is,  $1/S$ represents
a direct analog of the power exponent $n$ in the self-similar models of 
VB94, Uzdensky et al. 2002a, Low \& Lou (1990), and Wolfson (1995) (and 
the power exponent~$p$ in LBB94). Thus, the last equation can be 
interpreted as a statement that $n=1/S\sim\Delta\theta_2 \rightarrow 0$ 
as a thin current layer forms, in agreement with LBB94 and Wolfson (1995). 
This shows that a current sheet formation is intrinsically related to the 
increase in~$S$, and hence can be attributed to a growing difference in the 
expansion ratios of neighboring field lines with different twist angles, as 
discussed at the end of~\S~\ref{sec-intro}.


\section{Analysis and Interpretation of the Results}
\label{sec-analysis}

Thus we managed to reduce the problem to a system of two 
equations, namely, equation~(\ref{eq-5-C_S_Deltatheta2}) 
and equation~(\ref{eq-5-twist_constraint}). The latter
can be rewritten as 
\begin{equation}
{1\over{S\Delta\theta_2}} {{d\Delta\theta}\over{dx}}-
{{\Delta\theta}\over{\Delta\theta_2}} = -u(x).
\label{eq-6-main-x}
\end{equation}

Here 
\begin{equation}
x\equiv \log \Psi, 
\end{equation}
and
\begin{equation}
u(x) \equiv {{\Delta\Phi(x)}\over{\Delta\Phi_c}},
\end{equation}
where
\begin{equation}
\Delta\Phi_c \equiv {{\kappa a_1}\over C}=
{{2\kappa a_2}\over{\sin^2\theta_{\rm ap}}} = {\rm const}=O(1)     
\end{equation}
is some critical twist angle. For example, for $\kappa=1$, $a_2=1$, 
and~$\sin\theta_{\rm ap}\simeq 1$, we get $\Delta\Phi_c\simeq 2$~rad.
As we shall see below, the behavior of the system below and above 
$\Delta\Phi_c$ greatly differs. 

It is convenient to define a new independent variable 
\begin{equation}
t=\log r_{\rm ap}(\Psi). 
\label{eq-6-def-t}
\end{equation}
Then, 
\begin{equation}
S=-dt/dx,
\label{eq-6-S=-dt/dx}
\end{equation}
and equation~(\ref{eq-6-main-x}) becomes
\begin{equation}
{{\Delta\theta(t)}\over{\Delta\theta(t+1)}}
\left(1-{{d\log\Delta\theta}\over{dt}} \right)= u[x(t)]\, .
\label{eq-6-main-t}
\end{equation}

At this point we would like to pause in order to verify the validity 
of two steps in our program, namely the derivation of equations~(\ref
{eq-4-force_balance_theta}) and~(\ref{eq-5-radial_force_balance-1}) in
\S~\ref{sec-theta-balance} and~\S~\ref{sec-radial-balance}, respectively.
The validity of both these steps was based on a single assumption that,
at the apex $\theta=\theta_{\rm ap}$ the first term in the expression 
for the toroidal current density, $j_\phi=(1/r)(\partial/\partial r)
(rB_\theta)-(1/r)(\partial B_r/\partial\theta)$, is much smaller than 
the second term (in the limit $\Delta\theta\ll 1$ that we consider here).
At this stage in our analysis we now have at our disposal all the tools
necessary to estimate the magnitude of these two terms. 
Indeed, using equations~(\ref{eq-2-Btheta}), (\ref{eq-5-S<<1}), 
and~(\ref{eq-5-C_S_Deltatheta2}), the size of the first term can
be evaluated as
\begin{equation}
\biggl|{\partial\over{\partial r}}\Bigl(rB_\theta\Bigr)\biggr|_{\rm apex}
\sim \biggl|{{\partial^2 \Psi}\over{\partial r^2}}\biggr|_{\rm apex}\sim
{\Psi\over{r^2_{\rm ap}(\Psi)}}\, \biggl[ O\left({1\over S}\right) +
O\left(\left|{{d\Delta\theta_2}\over{dt}}\right|\right) \biggr] \ll
{\Psi\over{r^2_{\rm ap}(\Psi)}}\, .
\end{equation}

At the same time, using equations~(\ref{eq-2-Br}), and~(\ref
{eq-5-d2Psi/dtheta2}), the second term, $(1/r)(\partial B_r/
\partial\theta)$, can be estimated at $\theta=\theta_{\rm ap}$
as
\begin{equation}
\biggl|{1\over r}\, {{\partial B_r}\over{\partial\theta}} \biggr|_{\rm apex}
\sim {1\over r^2}\, \biggl|{{\partial^2\Psi}\over{\partial\theta^2}}
\biggr|_{\rm apex} \sim {\Psi\over{r^2_{\rm ap}(\Psi)}}\, 
{1\over{\Delta\theta_2}} \gg {\Psi\over{r^2_{\rm ap}(\Psi)}}\, .
\end{equation}
Thus, we see that the approximation we used is indeed well justified
in the small-$\Delta\theta$ limit.

We now turn to the application of our results to several particular 
situations.


\subsection{The Constant-$u$ Solution}
\label{subsec-const-u}

The {\it non-local} nature of equation~(\ref{eq-6-main-t}) 
makes it very difficult to analyze. The situation is complicated 
even further by the fact that the dependence $x(t)$ and, hence,
the right hand side of equation~(\ref{eq-6-main-t}) are not 
explicitly known. However, an exact solution of this equation 
can be found in the case of a disk rotating as a solid body, 
$\Delta\Phi(\Psi)={\rm const}$. This may be, in fact, a good 
approximation for the outer parts of a Keplerian disk, $r\gg 
r_{\rm co}$ (where $r_{\rm co}$ is the corotation radius),
where $\Delta\Omega(\Psi)=\Omega_{\rm disk}(\Psi)-\Omega_* 
\approx -\Omega_*={\rm const}$. In this {\it uniformly-rotating} 
disk model, one thus has $u(x)={\rm const}$. The solution that 
satisfies the boundary condition $\Delta\theta(t_0)=\Delta\theta_0$ 
is
\begin{equation}
\Delta\theta(t)=\Delta\theta_0 e^{(1-\xi)(t-t_0)},
\label{eq-6.1-deltatheta_t}
\end{equation}
and the constant $\xi$ is uniquely determined by $u$ via 
\begin{equation}
u=\xi e^{(\xi-1)}.
\label{eq-6.1-def-xi}
\end{equation}

Next, using equation~(\ref{eq-5-C_S_Deltatheta2}), we find
$S=S_0 \exp[(\xi-1)(t-t_0)]$, where $S_0\equiv S(t_0)=
u/C\xi\Delta\theta_0$. Then, using equation~(\ref{eq-6-S=-dt/dx}),
we can calculate~$x(t)$,
\begin{equation}
x=x_0+{1\over{\xi-1}}{1\over{S_0}}
\left( e^{(1-\xi)(t-t_0)}-1 \right),
\label{eq-6.1-x_t}
\end{equation}
and thus determine $\Delta\theta$ as a function of the 
poloidal flux:
\begin{equation}
\Delta\theta(x)=\Delta\theta_0+{u\over{C\xi}} (\xi-1) (x-x_0)=
\Delta\theta_0+{{\xi-1}\over C} e^{\xi-1} \log{\Psi\over{\Psi_0}}. 
\label{eq-6.1-deltatheta_x}
\end{equation}

For each $u$ there is a unique solution $\xi$, such that
$\xi(u>1)>1$ and $\xi(u<1)<1$. This means that, if $u>1$
(that is, $\Delta\Phi>\Delta\Phi_c$), then $\Delta\theta$ 
decreases outward, and if $u<1$ ($\Delta\Phi>\Delta\Phi_c$), 
then $\Delta\theta$ increases outward. The special case $u=1$
corresponds to $\xi=1$ and $\Delta\theta={\rm const}$ and thus 
describes the approach to the singularity $\Delta\Phi_c$ in the 
self-similar case discussed by VB94 and Uzdensky et al.~(2002a).

Notice that in the case $u<1$ ($\xi<1$), $\Delta\theta$ increases outward 
(with decreasing $\Psi$) until our model's main assumption $\Delta\theta 
\ll 1$ breaks down. This condition defines the outer boundary 
$\Psi_{\rm min}$ of applicability of the present model, which 
can be estimated roughly as the value of $\Psi$ for which 
$\Delta\theta \simeq 1$:
\begin{equation}
\Psi_{\rm min}(\xi) \simeq \Psi_0 
\exp \left[-{C\over{1-\xi}} e^{1-\xi}\right].
\end{equation}
We see that the range of validity of our model widens
($\Psi_{\rm min}$ decreases) as $\xi\rightarrow 1$.

In the opposite case $u>1$ ($\xi>1$)
there exists a certain critical field line,
\begin{equation}
\Psi_c=\Psi_0 \exp \left[-{{C\Delta\theta_0}\over{\xi-1}} e^{1-\xi}\right],
\end{equation}
on which $\Delta\theta$ vanishes while $r_{\rm ap}$ becomes equal 
to infinity. If $u$ (and hence~$\xi$) just barely exceeds~1, then 
this point is very far away: $\Psi_c \ll \Psi_0$ and $r_0(\Psi_c) 
\gg r_0(\Psi_0)$. As $u$ and $\xi$ increase, $\Psi_c$ also increases 
and the critical field line moves inward.

We interpret the vanishing of $\Delta\theta(\Psi_c)$ as a partial 
field-line opening, with $\Psi_c$ marking the boundary between 
open and closed field lines. Our model describes only closed 
field lines and thus is valid only for $\Psi>\Psi_c$. For any 
given field line $\Psi>\Psi_c$, $r_{\rm ap}(\Psi)$ is still 
finite, and, since $r_{\rm ap}(\Psi_c)=\infty$, the field line 
$\hat{\Psi}$ with $r_{\rm ap}(\hat{\Psi})=er_{\rm ap}(\Psi)$
still lies between~$\Psi$ and~$\Psi_c$. We can see then that
the analysis given in the previous sections will still be valid
for any field line $\Psi>\Psi_c$, no matter how close to~$\Psi_c$.
For open field lines, $\Psi<\Psi_c$, our treatment is, of course,
no longer valid. However, such an elaborate analysis is not needed 
for these field lines because their physics is very simple: the
open field ($\Psi<\Psi_c$) is potential, $F(\Psi)\equiv 0$, and
mostly radial.

We would like to remark that expression~(\ref{eq-6.1-deltatheta_x})
describes the behavior of the {\it angular width} of the current 
concentration region as a function of~$\Psi$. We can also see whether 
the {\it physical width} of this region (which probably is of greater
interest, as far as reconnection is concerned) decreases as~$\Delta\theta 
\rightarrow 0$. From equation~(\ref{eq-6.1-x_t}) we have
\begin{equation}
{{r_{\rm ap}(x)}\over{r_{\rm ap}(x_0)}}=
\left[S_0(\xi-1)(x-x_0)+1 \right]^{1\over{1-\xi}},
\end{equation}
and so
\begin{equation}
r_{\rm ap}(\Psi)\, \Delta\theta(\Psi)\, \propto \, 
[\Delta\theta(\Psi)]^{{2-\xi}\over{1-\xi}}.
\label{eq-6.1-r_deltatheta}
\end{equation}

We see that $r_{\rm ap}\Delta\theta \rightarrow 0$ with $\Delta\theta$ 
if $\xi>2$, i.e., if $u>2e\approx 5.4$. Thus, we envision the following 
scenario. As the twist is increased beyond $u=1$, there will be a line 
$\Psi_c$ beyond which the field will be open: 
$r_{\rm ap}(\Psi\rightarrow\Psi_c)\rightarrow \infty$.
As the twist is increased even further, more and more flux becomes open. 
At the same time, as~$u$ is increased beyond~$2e$, then, according to 
the power law~(\ref{eq-6.1-r_deltatheta}), not only the angular, but 
also the actual physical width of the current concentration region, 
$r_{\rm ap}\Delta\theta$, shrinks to zero as one takes the limit 
$\Psi\rightarrow\Psi_c$. This suggests that increasing the twist 
angle well above the critical value can not only result in a greater 
amount of open flux, but also can lead to a narrower current layer, 
thus facilitating magnetic reconnection.

We would like to emphasize that, while our analysis predicts 
a finite-time partial field opening, it does not, strictly 
speaking, predict a finite-time current-sheet formation in 
a true sense. Indeed, at any given $\Delta\Phi>\Delta\Phi_c$, 
and for any given still-closed field line~$\Psi>\Psi_c(\Delta\Phi)$, 
the current concentration region has a non-zero angular thickness
$\Delta\theta(\Psi,t)$  (which, however, tends to zero as 
$\Psi\rightarrow\Psi_c$). If one fixes a field line~$\Psi$, 
and monitors how the angular thickness $\Delta\theta(\Psi,t)$ 
changes with time, then one finds that it does indeed go to 
zero in finite time at the moment of opening of this field 
line. If, however, one looks at any given {\it finite radius}~$r$ 
(instead of fixing a field line~$\Psi$), then the situation is very
different. Since $r_{\rm ap}(\Psi_c)=\infty$, there is always a 
closed field line $\Psi>\Psi_c$ with $r_{\rm ap}(\Psi)=r$; thus, 
the angular thickness $\Delta\theta(r,t)$ stays finite at the 
moment of first field-line opening (when $\Delta\Phi=\Delta\Phi_c$). 
As the twisting continues, $\Delta\theta(r,t)$ will probably decrease 
and may in fact go to zero asymptotically as $t\rightarrow\infty$.
Thus, one may say that even though one has a finite-time partial 
field-line opening, there is no true finite-time current sheet 
formation. This is indeed a very important distinction. Of course, 
on a practical note, what is important here is the fact that the 
current layer becomes very thin: at some point during the thinning 
process finite resistivity may come into play, leading to field-line 
reconnection.

Finally, we need to point out that, when partial field 
opening takes place, some parts of an opening field line 
move very rapidly (e.g., the velocity of the apex approaches 
infinity at the point of opening). We have to make it clear 
that these infinite velocities are an artifact of the 
equilibrium assumption. They should be understood only
in the sense that they become very large compared with
the footpoint rotation velocity. In reality, the expansion
velocity will be limited, most likely, by finite inertia of 
the plasma, i.e., by the Alfv{\'e}n speed. 
Another concern is the assumption that the suggested sequence of 
partially-open field configurations is continuous. It is actually 
not completely clear whether this always has to be so. In reality,
it may be possible that, upon reaching a critical twist, a quasi-static 
evolution along a sequence of force-free fields will end abruptly by 
opening up a finite portion of the flux.


\subsection{Solution in the vicinity of $\Delta\Phi_c$}
\label{subsec-critical-twist}

Another important advance in our understanding 
of equation~(\ref{eq-6-main-t}) can be made in 
a situation where $\Delta\Phi(\Psi)$ is nonuniform 
and passes through the critical twist angle $\Delta\Phi_c$ 
(for example, as one moves outward, i.e., as $\Psi$ is decreased).

Let us consider a close vicinity of the point $x_0 \equiv \log\Psi_0$
where $u(x_0)=1$. In this region we can Taylor-expand $u(x)$ and keep
only the linear term:
\begin{equation}
u(x)=1-\eta (x-x_0), \qquad {\rm where}\ \eta=-{{du}\over{dx}}(x_0)>0.
\label{eq-6.2-u(x)}
\end{equation}

Let us define two functions,
\begin{equation}
\epsilon(t)\equiv 
{{\Delta\theta(t+1)-\Delta\theta(t)}\over{\Delta\theta(t)}},
\label{eq-6.2-def-epsilon}
\end{equation}
and
\begin{equation}
\sigma(t)\equiv {{d\log\Delta\theta(t)}\over{dt}}\, ,
\label{eq-6.2-def-sigma}
\end{equation}
and consider a region around $t_0=t(x_0)$ corresponding 
to a close vicinity of $x_0$, such that $|u-1|\ll 1$ within 
this region. We anticipate that $\Delta\theta$ varies relatively 
little so that $|\epsilon(t)| \ll 1$ everywhere in this region.

Then equation~(\ref{eq-6-main-t}) becomes 
\begin{equation}
(1-\epsilon)(1-\sigma)=u\approx 1,
\end{equation}
and therefore $|\sigma(t)| \ll 1$, which gives us a relationship
between~$\epsilon(t)$ and~$\sigma(t)$:
\begin{equation}
\epsilon(t)+\sigma(t)=1-u(t) \ll 1.
\label{eq-6.2-epsilon+sigma}
\end{equation}

A second relationship follows from the definitions of
$\epsilon(t)$ and~$\sigma(t)$:
\begin{equation}
\epsilon(t) \simeq \int\limits_t^{t+1} \sigma(\hat{t}) d\hat{t},
\label{eq-6.2-epsilon=integral_sigma}
\end{equation}
where we have neglected higher-order corrections in~$\sigma(t)$.
Using this expression we can differentiate equation~(\ref
{eq-6.2-epsilon+sigma}) and get
\begin{equation}
\sigma'(t)+\sigma(t+1)-\sigma(t) = -u'(t)\, ,
\end{equation}
where prime denotes the derivative with respect to~$t$.

Now let us Taylor-expand $\sigma(t+1)$ around the point~$t$ and
keep only the first two terms:
\begin{equation}
\sigma(t+1) \approx \sigma(t)+\sigma'(t) \, .
\label{eq-6.2-Taylor_expansion}
\end{equation}
Then we can write
\begin{equation}
\sigma'(t)=-{{u'(t)}\over 2} \, .
\label{eq-6.2-sigma'-1}
\end{equation}

We now need to make a little digression in order to discuss 
the validity of this step, that is of keeping only the first 
two terms in the expansion of $\sigma(t+1)$. In particular, 
we need to verify that we can safely neglect the next term in 
the Taylor expansion, $0.5\, \sigma''(t)$. This term needs to 
be compared with the last term that we kept in the expansion, 
that is with~$\sigma'(t)$, which, according to equation~(\ref
{eq-6.2-sigma'-1}), can be rewritten as 
\begin{equation}
\sigma'(t)=-{{u'(t)}\over 2}={1\over{2S}}\, {du(x)\over{dx}}=
{C\over 2}\, {du(x)\over{dx}}\, \Delta\theta(t+1)\, ,
\label{eq-6.2-sigma'-2}
\end{equation} 
where we made use of equations~(\ref{eq-6-S=-dt/dx}) 
and~(\ref{eq-5-C_S_Deltatheta2}). At the same time, 
the first neglected term is
$$
{{\sigma''(t)}\over 2} = -{{C^2}\over 4}\, {{d^2u(x)}\over{dx^2}}\, 
\Delta\theta^2(t+1)\, + {C\over 4}\, {du(x)\over{dx}}\,
\Delta\theta(t+1)\, \sigma(t+1)\, ,
$$
where we again used equations~(\ref{eq-5-C_S_Deltatheta2}) and
(\ref{eq-6-S=-dt/dx}), as well as equation~(\ref{eq-6.2-def-sigma}). 
The second term in this expression is manifestly much smaller than 
(\ref{eq-6.2-sigma'-2}) because we here are considering a region 
where $|\sigma(t)|\ll 1$. Thus we see that our approximation (\ref
{eq-6.2-Taylor_expansion}) is valid if
\begin{equation}
\biggl|{{du(x)}\over{dx}}\biggr| \simeq \eta \gg 
\biggl|{{d^2u(x)}\over{dx^2}}\biggr|\, \Delta\theta(t+1)\, .
\end{equation}
The function $u(x)$ is the prescribed twist angle as a function of flux;
generally, it varies on a scale that is finite in~$\Psi$ and, hence,
in~$x$. This means that both $du/dx$ and $d^2u/dx^2$ are of order 1; 
therefore, since we work in the regime $\Delta\theta\ll 1$, the above 
condition is easily satisfied and thus our truncation of the Taylor 
expansion~(\ref{eq-6.2-Taylor_expansion}) is valid.

We now continue with our analysis. 
Since we consider a region where $\Delta\theta$ is almost constant, we
can, to the lowest order in $\sigma$, substitute $\Delta\theta(t+1)$
in equation~(\ref{eq-6.2-sigma'-2}) by 
$\Delta\theta_0\equiv\Delta\theta(t_0)$, where $t_0$ is defined by 
$x(t_0)=x_0$. In a similar vain, $du/dx$ also varies very little in 
the region under consideration, so, using equation~(\ref{eq-6.2-u(x)}) 
we can substitute it by $-\eta$. Then we can integrate equation~(\ref
{eq-6.2-sigma'-2}) to get
\begin{equation}
\sigma(t)=\sigma_0 -{{C\eta}\over 2} \Delta\theta_0 (t-t_0) +O(\sigma^2),
\end{equation}
where $\sigma_0\equiv\sigma(t_0)$.

Combining this expression with the definition of~$\sigma(t)$, 
we get the following expression for $\Delta\theta(t)$:
\begin{eqnarray}
\Delta\theta(t) &=& \Delta\theta_0\,  
\exp\biggl[\sigma_0(t-t_0)\biggr] \, \cdot\, 
\exp\biggl[-{{C\eta}\over 4}\Delta\theta_0 (t-t_0)^2 \biggr] \nonumber \\ 
&\simeq& \Delta\theta_0 \left( 1 + \sigma_0 (t-t_0) -
{{C\eta}\over 4} \Delta\theta_0 (t-t_0)^2 \right).
\end{eqnarray}

Then, from equation~(\ref{eq-6.2-epsilon=integral_sigma}) it follows that
\begin{equation}
\epsilon(t)=\sigma_0 - {{C\eta}\over 4}\Delta\theta_0 [1+2(t-t_0)],
\end{equation}
and from equation~(\ref{eq-6.2-epsilon+sigma}) we get
\begin{equation}
\sigma_0 \equiv \sigma(t_0) = {C\eta\over 8} \Delta\theta_0 > 0.
\end{equation}

Thus, we can write
\begin{equation}
\Delta\theta(t)=\Delta\theta_0 \Bigl[ 1 + \sigma_0 (t-t_0)
[1-2(t-t_0)] \Bigr].
\end{equation}

We see that $\Delta\theta(t)$ behaves regularly at~$t=t_0$; 
for~$t>t_0$, it first increases, reaches a maximum at $t=t_0+1/4$ 
(i.e., $x-x_0=-C\Delta\theta_0/4 \ll 1$), and then starts to 
decrease. If $u>1$ for all $x<x_0$, then, guided by our constant-$u$ 
solution, we can predict that $\Delta\theta$ at some point will reach 
zero, implying a partial field-line opening.


\subsection{Locally enhanced twist}
\label{subsec-numerical}

As a last example, we present the results of numerical solution
of the system (\ref{eq-6-main-t})--(\ref{eq-6-S=-dt/dx}) for the 
case of a locally enhanced~$u(x)$.

\begin{figure}[tbp]
\begin{center}
\epsfig{file=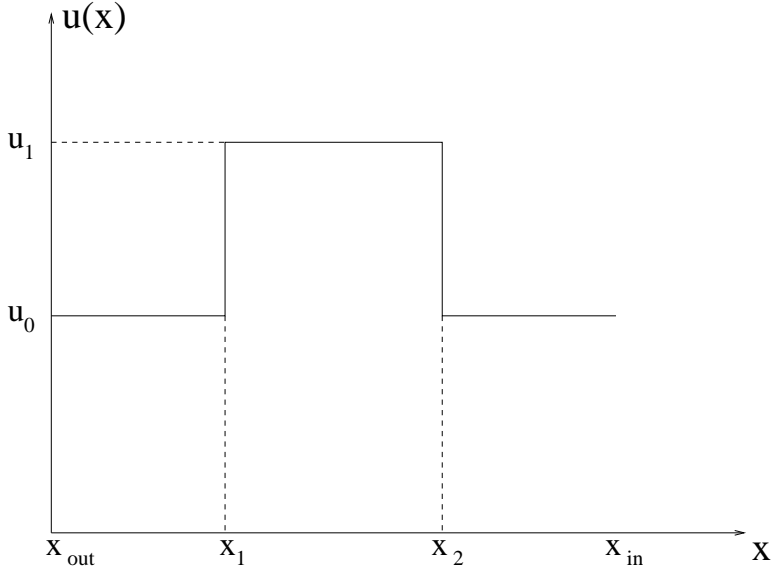,width=4 in}
\caption{The $u(x)$ profile investigated in \S~\ref{subsec-numerical}.
\label{fig-u-of-x}}
\end{center}
\end{figure}

We consider an interval $x\in [x_{\rm out},x_{\rm in}]$
corresponding to $\Psi\in [\Psi_{\rm out},\Psi_{\rm in}]$.
In order to be able to make a connection with the constant-$u$ 
analytical solution, we take the function~$u(x)$ to consist of 
three pieces (see Figure~\ref{fig-u-of-x}):\\
$u(x)=u_0<1, x_{\rm out} \leq x \leq x_1$; \\
$u(x)=u_1>1, x_1<x<x_2$; \\
and \\
$u(x)=u_0, x_2 \leq x \leq x_{\rm in}$.

Here, $u_0={\rm const}<1$ is some background value, which 
we keep fixed, and $\Delta u = u_1-u_0={\rm const}$ is the 
localized enhancement. We study a family of solutions of the 
system (\ref{eq-6-main-t})--(\ref{eq-6-S=-dt/dx}) with a fixed 
value of $u_0$, $u_0=0.655..$ [which corresponds to $\xi_0\equiv 
\xi(u_0) =0.8$], but with several different values of~$u_1$. 
We start with $u_1=1.2$, and then gradually increase it until 
the sequence terminates at $u_1 \simeq 1.8$. In order for us 
to be able to make a meaningful comparison of the solutions 
with different values of $u_1$, we fix the value of $\Delta\theta$ on 
the innermost field line, $\Delta\theta(x_{\rm in})=\Delta\theta_0$.

Now, based on our experience with constant-$u$ solutions,
we can naively expect the resulting function $\Delta\theta(x)$ 
to be roughly a piece-wise linear function, with $\Delta\theta(x)$
decreasing linearly in the intervals $x_{\rm out}<x<x_1$ and 
$x_2<x<x_{\rm in}$ [with the slope $u_0(\xi_0-1)/C\xi_0$], 
and increasing linearly in the interval $x_1<x<x_2$ [with 
the slope $u_1(\xi_1-1)/C\xi_1$, where $\xi_1 \equiv \xi(u_1)>1$]. 
We thus expect $\Delta\theta$ to have a maximum at $x_2$, 
$\Delta\theta^{\rm est}_{\rm max} = \Delta\theta(x_2)=
\Delta\theta_0+ u_0(\xi_0-1)(x_2-x_{\rm in})/C\xi_0$, and  
a minimum at $x_1$, $\Delta\theta^{\rm est}_{\rm min}=\Delta\theta(x_1)
=\Delta\theta^{\rm est}_{\rm max}-u_1(\xi_1-1)(x_2-x_1)/C\xi_1$. 

Note that, when $u_1$ reaches a certain value $u_{1,\rm max}>1$ 
[and, correspondingly, $\xi_{1,\rm max} \equiv \xi(u_{1,\rm max})>1$] 
such that
\begin{equation}
(\xi_{1,\rm max}-1) \, \exp (\xi_{1,\rm max}-1) = 
{{C\Delta\theta_{\rm max}}\over{x_2-x_1}}\, , 
\label{eq-6.3-u_1max}
\end{equation}
then $\Delta\theta^{\rm est}_{\rm min}$ becomes zero.  
We interpret this as the point in the sequence of equilibria 
at which a {\it partial field-line opening} occurs at 
$\Psi=\Psi_1$. Note that $u_{1,\rm max}$, i.e., the 
maximum allowable value of~$u_1$, depends inversely 
on the size $(x_1-x_2)$ of the enhanced-twist region.
We thus see that, if only a portion of field lines
(those between $\Psi_1$ and $\Psi_2$) are participating 
in twisting and expansion, then the thinning of the current 
layer, and the corresponding partial field-line opening,
is achieved not at $u=1$, but at a somewhat higher value.

When performing the actual computation, we advance along~$t$ from 
$t_{\rm out}$ (corresponding to the outermost field line~$\Psi_{\rm 
out}$) inward. We first pick a guess for the initial value of 
$\Delta\theta$ at the outer boundary, $\Delta\theta_{\rm out}$,
and then use our const-$u$ solution to prescribe the function
$\Delta\theta(t)$ and hence $x(t)$ in the {\it initial interval}
of unit length, $t\in [t_{\rm out}-1,t_{\rm out}]$. After that,
we proceed to integrate the two differential equations:
\begin{eqnarray}
{{d\Delta\theta}\over{dt}} = \Delta\theta(t)-u[x(t)] \Delta\theta(t+1), \\
{{dx}\over{dt}} = -{1\over{S(t)}} = -C\Delta\theta(t+1).
\end{eqnarray}

We stop when the innermost field line of the domain, $x=x_{\rm in}$, 
$\Psi=\Psi_{\rm in}$, is reached and compare the computed value 
of $\Delta\theta$ at this point with the prescribed value 
$\Delta\theta_{\rm in}=\Delta\theta_0$. We then iterate with respect to 
$\Delta\theta_{\rm out}$, i.e., we change $\Delta\theta_{\rm out}$
and repeat the procedure until $\Delta\theta(x_{\rm in})$ finds 
itself within a small vicinity (which we usually take to be~1\%) 
of the desired value~$\Delta\theta_0$.

\begin{figure}[tbp]
\begin{center}
\epsfig{file=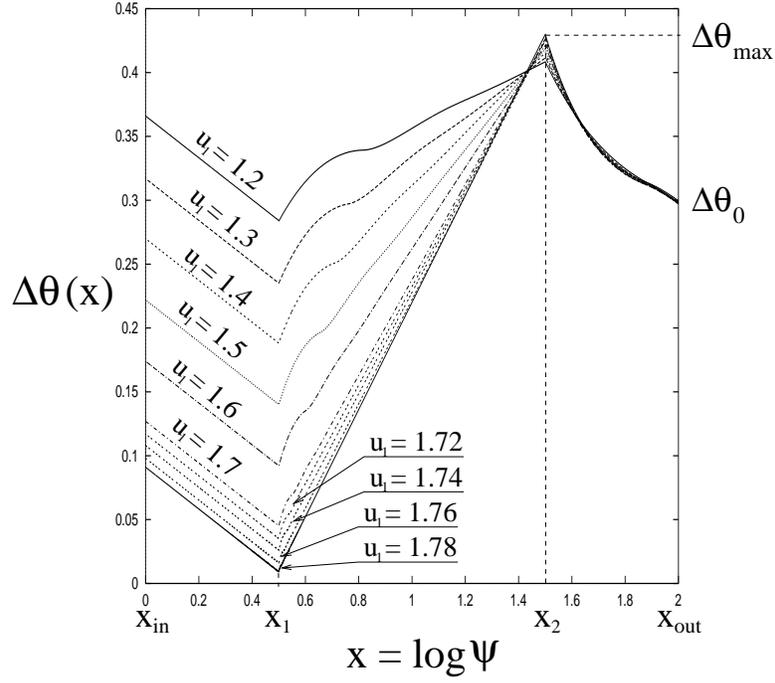,width=4 in}
\caption{Angular thickness $\Delta\theta$ as a function 
of~$x\equiv \log \Psi$ for a range of values of~$u_1$ between 
$u_1=1.2$ and~$u_1=1.8$.
\label{fig-deltatheta}}
\end{center}
\end{figure}

In Figure~\ref{fig-deltatheta} we present the results of 
our computations for $\Delta\theta_0=0.3$, $x_{\rm out}=0.0$, 
$x_1=0.5$, $x_2=1.5$, $x_{\rm in}=2.0$, $C=1$, $u_0=0.655$, and 
$u_1$ between 1.2 and 1.8. We indeed see that the obtained solutions
$\Delta\theta(x)$ are in good agreement with the analytically-predicted
picture described above. In particular, we find $\Delta\theta_{\rm max} 
\simeq 0.43$, and $u_{1,\rm max}\simeq 1.8$. This is very close to the 
naive prediction~(\ref{eq-6.3-u_1max}), which gives $\Delta\theta_{\rm 
max}=0.382$, $\xi_{1,\rm max}=1.287$, and $u^{\rm est}_{1,\rm max}=1.71$ 
for the set of parameters that we chose. And, if we use the actual value 
$\Delta\theta_{\rm max}=0.43$ in equation~(\ref{eq-6.3-u_1max}), we then 
get $\xi_{1,{\rm max}}=1.314$, and $u^{\rm est}_{1,\rm max}=1.80$.

\begin{figure}[tbp]
\begin{center}
\epsfig{file=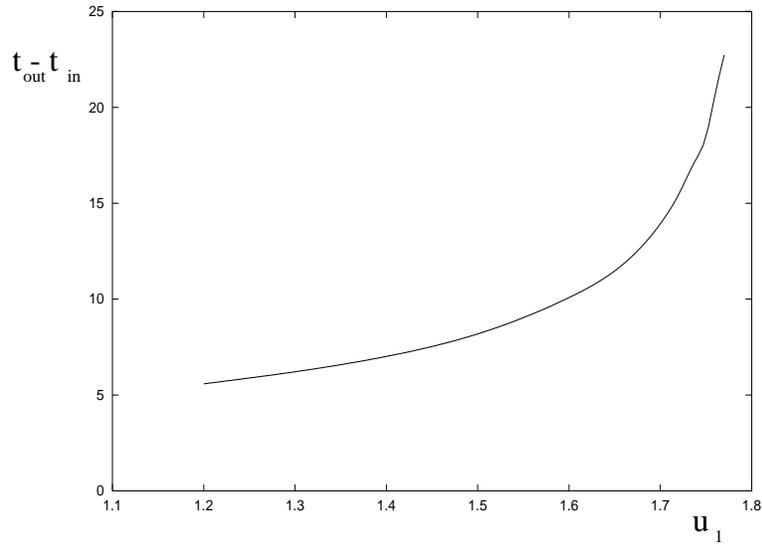,width=4 in}
\caption{Natural logarithm of the ratio of $r_{\rm ap}$ of the outermost 
field line to $r_{\rm ap}$ of the innermost field line,
$t_{\rm out}-t_{\rm in} \equiv \log[r_{\rm ap}(\Psi_{\rm out})/
r_{\rm ap}(\Psi_{\rm in})]$, as a function of the twist~$u_1$.
\label{fig-expansion}}
\end{center}
\end{figure}

At the same time we observe that, as $u_{1,\rm max}$ is approached, 
the apexes of the field lines move out to infinity explosively fast,
signaling a partial field-line opening at a finite twist.
This opening can be illustrated by the behavior of the
natural logarithm of the ratio of $r_{\rm ap}$ of the outermost 
field line to $r_{\rm ap}$ of the innermost field line,
i.e., by $t_{\rm out}-t_{\rm in} \equiv \log[r_{\rm ap}(\Psi_{\rm out})/
r_{\rm ap}(\Psi_{\rm in})]$. Figure~\ref{fig-expansion} shows this ratio
as a function of $u_1$. We thus see an unambiguous evidence 
that the thinning of the current layer occurs simultaneously
with the partial field-line opening, in accord with our expectations
discussed in \S~\ref{sec-intro}.

\section{Conclusions} 
\label{sec-conclusions}

In this paper we addressed the issue of the formation of thin 
current structures in the magnetosphere of an accretion disk 
in response to the field-line twisting. We presented a simple 
analytical model that illustrates the asymptotic behavior of a 
force-free axisymmetric magnetic field above a thin conducting
disk in the limit when such structures are about to form. We 
showed that there exist a finite critical twist angle, 
$\Delta\Phi_c$, such that the expansion of the poloidal field 
lines accelerates dramatically as $\Delta\Phi_c$ is exceeded.
At the same time, in the case of a locally-enhanced twist,%
\footnote{
An important example of the locally-enhanced-twist case is 
a Keplerian disk whose inner edge is significantly closer to 
the central star than the corotation radius. In particular, if 
$r_{\rm inner}<0.63\, r_{\rm co}$, then the relative star--disk 
angular velocity, $\Delta\Omega(r)=\Omega_{\rm disk}(r)-\Omega_*$, 
reaches higher values in the localized inner region ($r_{\rm inner}<
r<r_{\rm co}$) than at large distances ($r>r_{\rm co}$).}
a radially-extended thin conical current layer forms. 
Simultaneously, the field configuration approaches a 
partially-open state in finite time, with the apexes 
of a portion of the field lines going out to infinity. 

The presence of a thin current sheet is usually considered
a pre-requisite for reconnection of magnetic field lines.
It is also believed that reconnection will occur whenever
a thin current layer is formed. We therefore believe that 
magnetic field lines that have become effectively open as 
a result of their twist-driven inflation may subsequently 
reconnect via the thin current layer that has been formed 
simultaneously with their opening-up. This conclusion leads 
to important consequences regarding the time evolution of 
magnetospheres of magnetically-linked star--disk systems. 
In particular, we believe that, once the magnetic link between 
the star and the disk is broken by the partial field-line opening, 
a current layer will form along the separatrix between the stellar 
field lines and the disk field lines and, as a result, the link will 
be re-established through reconnection of these field lines. 
Subsequently, if reconnection process is sufficiently fast, 
the field lines will contract to a less-stressed configuration 
allowing a new cycle of twisting, inflation, opening, and reconnection
to begin, as was suggested by Aly \& Kuijpers (1990) and by VB94 and
illustrated in numerical simulations by Goodson et al. (1999). This 
quasi-periodic scenario is characterized by very rich physics with 
non-steady and violent behavior, perhaps not too different from that 
of the solar corona (e.g., Low 2001). It can provide avenues for 
understanding such phenomena as disk winds, time-variable, knotted jets, 
episodes of rapid accretion, and variable accretion torque on the central 
star.

One should note, however, that a true opening will be preceded 
by the plasma inertia becoming important. The inertial effects 
will tend to retard the expansion, since the pressure of the 
toroidal field pressure will have to work not only against the 
poloidal field tension, but also against the plasma inertia. 
Hence the inertial effects will in effect act against the 
toroidal field removal and will not help to form a current 
sheet. At some point during the expansion they will need to 
be taken into account and, even in the absence of reconnection, 
a true finite-time field opening will not happen. Instead, one 
will have a transition from the force-free regime into the inertial 
regime (wind regime), which would require solving the full set of 
MHD equations, although probably under some simplifying assumptions.

I am very grateful to Stanislav Boldyrev, Arieh K{\"o}nigl, and
Bob Rosner for very useful discussions and comments. I am also
grateful to the anonymous referee for his or her deep and insightful
criticism that was very beneficial for the paper. I would also
like to acknowledge the support by the ASCI/Alliances Center for 
Astrophysical Thermonuclear Flashes at the University of Chicago 
under DOE subcontract B341495 and by the NSF grant NSF-PHY99-07949.


\section*{REFERENCES}
\parindent 0 pt

Aly, J.~J. 1984, ApJ, 283, 349

Aly, J.~J., \& Kuijpers, J. 1990, A\&A, 227, 473

Aly, J.~J. 1995, ApJ, 439, L63

Barnes, C.~W., \& Sturrock, P.~A. 1972, ApJ, 174, 659

Bertout, C., Basri, G., \& Bouvier, J. 1988, ApJ,
330, 350

Ghosh, P., \& Lamb, F.~K. 1978, ApJ, 223, L83 (GL)

Ghosh, P., \& Lamb, F.~K. 1979a, ApJ, 232, 259 (GL)

Ghosh, P., \& Lamb, F.~K. 1979b, ApJ, 234, 296 (GL)

Goodson, A.~P., Winglee, R.~M., \& B{\"o}hm, K.-H. 1997,
ApJ, 489, 199
 
Goodson, A.~P., B{\"o}hm, K.-H., \& Winglee, R.~M. 1999,
ApJ, 524, 142
 
K{\"o}nigl, A. 1991, ApJ, 370, L39

Lamb, F. K. 1989, in Timing Neutron Stars, ed. H. {\"O}gelman \& 
E. P. J. van den Heuvel (Dordrecht: Kluwer), 649

Lovelace, R.~V.~E., Romanova, M.~M., \& Bisnovatyi-Kogan, G.~S.
1995, MNRAS, 275, 244 

Low, B.~C. 1986, ApJ, 307, 205

Low, B.~C., \& Lou, Y.~Q. 1990, ApJ, 352, 343

Low, B.~C. 1990, ARA\&A, 28, 491

Low, B.~C. 2001, JGR, 106, 25141

Lynden-Bell, D., \& Boily, C. 1994, MNRAS, 267, 146 (LBB94)

Miki{\'c}, Z., \& Linker, J.~A. 1994, ApJ, 430, 898

Parker, E.~N. 1979, Cosmical Magnetic Fields (Oxford: Oxford 
Univ. Press)

Patterson, J. 1994, PASP, 106, 209

Roumeliotis, G., Sturrock, P.~A., \& Antiochos S.~K. 1994;
ApJ, 423, 847

Uzdensky, D.~A., K{\"o}nigl, A., \& Litwin, C. 2002a,
ApJ, 565, 1191; also astro-ph/0011283

Uzdensky, D.~A., K{\"o}nigl, A., \& Litwin, C. 2002b,
ApJ, 565, 1205; also astro-ph/0011283

van~Ballegooijen, A.~A. 1994, Space Sci. Rev., 68, 299 (VB94)

Wang, Y.-M. 1987, A\&A, 183, 257

Wolfson, R., \& Low, B.~C. 1992, ApJ, 391, 353

Wolfson, R. 1995, ApJ, 443, 810

\end{document}